\documentclass[11pt]{article} 
\usepackage{amsmath} 
\usepackage{amsthm} 
\usepackage{amssymb}	
 \usepackage{indentfirst}
\usepackage{float}
\usepackage{titlesec}
\titlelabel{\thetitle. \quad}

\usepackage{authblk}
\usepackage{graphicx} 
\usepackage[dvips,letterpaper,margin=0.75in,bottom=0.5in]{geometry}
\usepackage[titletoc,toc,title]{appendix}
\usepackage{verbatim}
\usepackage{cite}

\newcommand{\matrixel}[3]{\left< #1 \vphantom{#2#3} \right|
 #2 \left| #3 \vphantom{#1#2} \right>} 

\begin{document}
\title{Two Photon Exchange in Impact Parameter Space  in the Relativistic Eikonal Approximation for Elastic $e-N^{\uparrow}$ Scattering}

\author{Tareq Alhalholy and Roman H{\"o}llwieser}
\affil{Department of Physics, New Mexico State University,
Las Cruces, NM 88003-0001, U.S.A.}
\date{\today}
\maketitle

\begin{abstract}

Using the relativistic Eikonal approximation, we study the one and two photon exchange amplitudes in elastic  electron-nucleon scattering for the case of transversely polarized nucleons with unpolarized electrons beam.  In our approach, we utilize the convolution theory of Fourier transforms and the transverse charge density in transverse momentum space to evaluate the one and two photon exchange Eikonal  amplitudes. The  results obtained for the $2 \gamma$ amplitude in impact parameter space  are compared to the corresponding $4-$D  case. We show that while the one and two  photon cross sections are azimuthally symmetric,  the interference term between them is azimuthally asymmetric, which is an indication of an azimuthal single spin asymmetry for  proton and neutron which can be attributed to the fact that the nucleon charge density is transversely (azimuthally) distorted in the transverse plane for transversely polarized nucleons. In addition, the calculations of the interference term for proton and neutron  show  agreement in sign and magnitude of the existence data and calculations for transverse target single spin asymmetry.
\end{abstract}

\titlepage



\section{Introduction}

\begin{figure}[H]
 \begin{center}
  \includegraphics[width=0.35\textwidth]{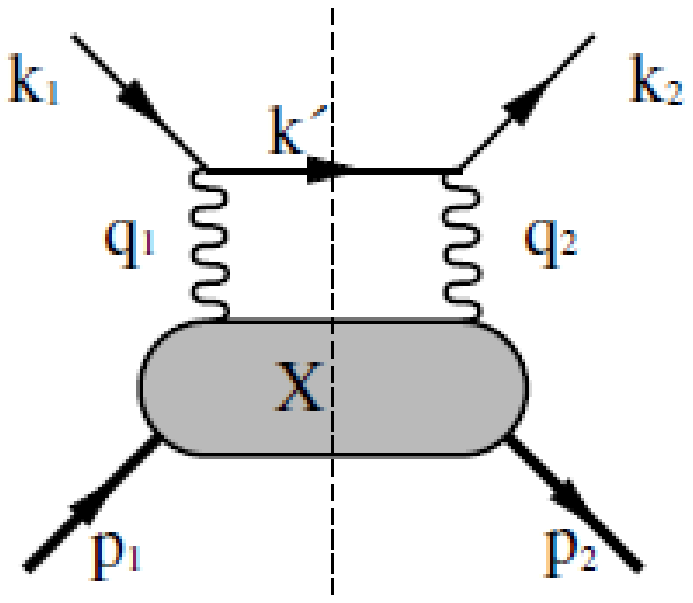}
   \includegraphics[width=0.45\textwidth]{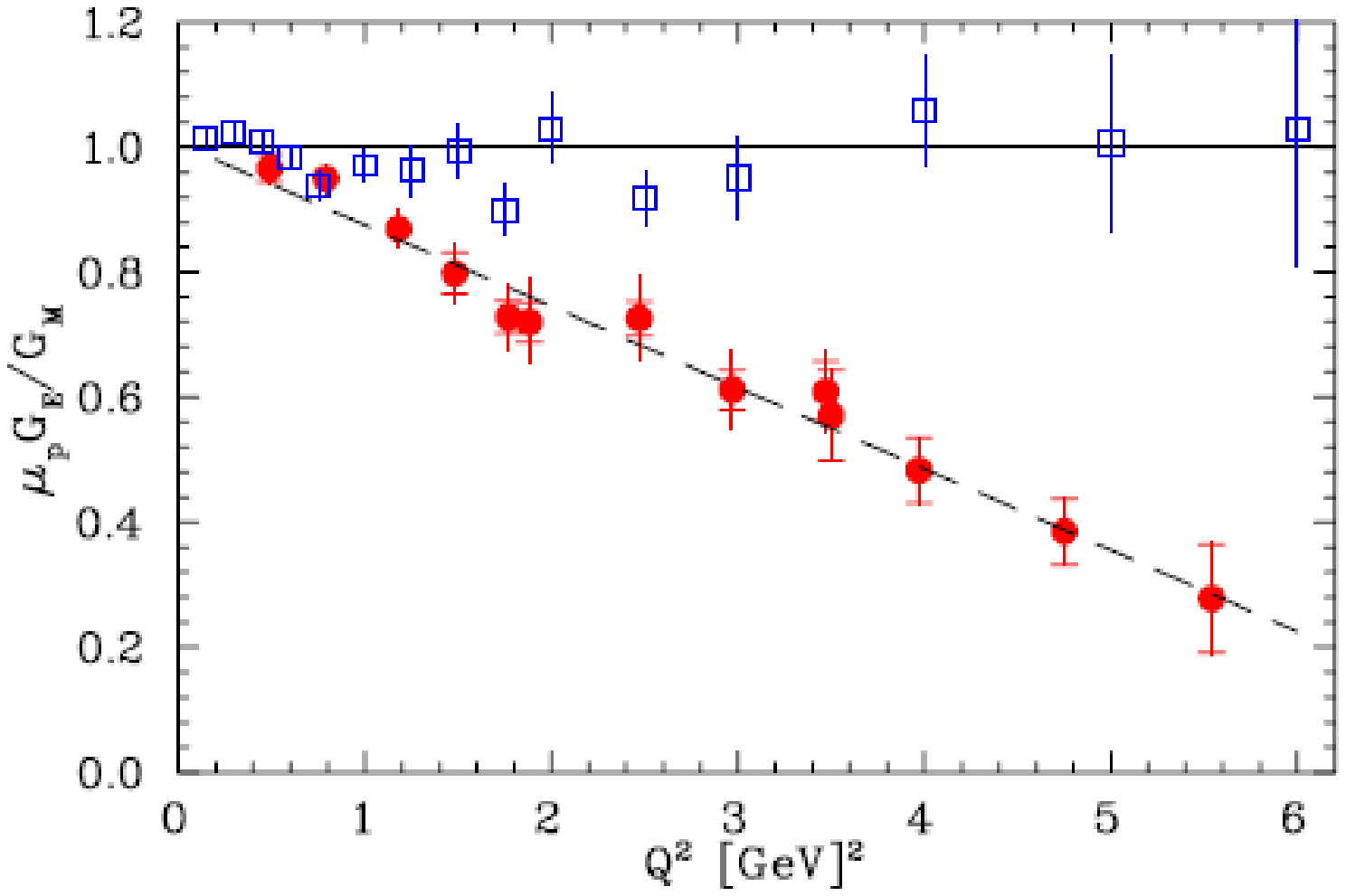}
 \end{center}
\caption{Left: two photon exchange box diagram. Right: Data of electromagnetic form factor ratio for proton, circles:  polarization transfer, squares: cross section measurements (Rosenbluth technique). }
 \label{fig:TwoPhEx}
\end{figure}

Two photon exchange has recently become an attractive subject due to it's importance in the calculations of different quantities  in elastic and deep inelastic scattering processes, for example single spin asymmetry is proportional to the interference of the one and two photon exchange amplitudes both in the elastic \cite{AfanaseveTPE1, Blunden2-2PE} and deep inelastic \cite{Metz1,Metz2} cases. It is also recently noted that the two photon exchange calculations are important to increase the precision in related quantities, for example the need for precision to resolve the  inconsistency in the measurement of the electromagnetic form factors ratio $G_E/G_M$  using  different methods of measurements~\cite{PartonicCalc,Qattan1,Blunden-2PE}, as shown in Figure.\ref{fig:TwoPhEx}.
The extension of the two photon exchange calculations of  the unpolarized case to polarized cases is important especially when dealing with quantities that are directly related to the $2 \gamma$ amplitude, such as single and double spin asymmetries. On the other hand, knowledge of  the impact parameter dependence of the $2 \gamma$ exchange process has the advantage of connecting  it with  different quantities in  impact parameter space , which is currently of prime importance  in studying the spin structure of the nucleon \cite{BurkardtGPDs,MillerRev} and in high energy elastic and deep inelastic scattering processes.

In this work, we study the one and two photon exchange amplitudes for  transversely polarized target in elastic $e-N^{\uparrow}$ scattering in impact parameter space utilizing the relativistic Eikonal approximation. The obtained results are compared to the conventional four dimensional case of the two photon exchange process, where a notable similarity exist between the two cases; for example, as a result of our calculations, we noticed (for proton and neutron) that  while the one and two  photon elastic cross sections are azimuthally symmetric,  the interference  of the corresponding elastic one and two photon exchange amplitudes in impact parameter space is azimuthally asymmetric which is an indication of a non zero azimuthal single spin asymmetry for both  proton and neutron. The calculations are done for GPD and Sachs parametrization of the nucleon's form factors for proton and neutron.  


\section{Nucleon Transverse Charge and Magnetization Densities}
 For a nucleon of mass $M_N$, transversely polarized  (with respect to it's momentum direction) in the $x$ direction, the charge density is no longer axially symmetric and the  distribution of partons in the transverse plane is given by
\begin{multline}
q(x,\mathbf{b_\perp})  =  \int \frac{d^2  q_\perp}{(2 \pi)^2} e^{-i \mathbf{b_\perp} \cdot \mathbf{ q_\perp}} \left[ H^q (x,0,-q^2_\perp) + i \frac{q_y}{2 M_N} E^q (x,0,-q^2_\perp) \right]
  = q(x, b_\perp) - \frac{1}{2 M_N} \frac{\partial}{\partial b_y} \mathcal{E} ^q (x,\mathbf{b_\perp}),
\label{eq:PartonDist}
\end{multline}
where
\begin{equation}
q(x,\mathbf{b_\perp}) = \int \frac{d^2 q_\perp}{(2 \pi)^2} H^q (x,0, -q^2_\perp) e^{-i \mathbf{b_\perp} \cdot \mathbf{q_ \perp}}.
\end{equation}
\begin{equation}
\mathcal{E} ^q (x,\mathbf{b_\perp}) = \int \frac{d^2 q_\perp}{(2 \pi)^2} E^q (x,-q^2_\perp) e^{-i \mathbf{b_\perp} \cdot \mathbf{q_ \perp}},
\end{equation}
here 
 $ {\bf b_\perp} = b_\perp \cos (\phi_{b_\perp}) \ \hat{e}_x + b_\perp \sin(\phi_{b_\perp})  \ \hat{e}_y $ \ is the transverse (impact parameter) vector, \  ${\bf q_\perp} = q_\perp \cos (\phi_{q_\perp}) \ \hat{e}_x + q_\perp \sin(\phi_{q_\perp}) \ \hat{e}_y$ \ is the transverse momentum transfer and  the nucleon spin is  transverse to the incident beam direction with spin vector $\mathbf{S} = \cos (\phi_s) \ \hat{e}_x + \sin(\phi_s) \ \hat{e}_y$.
Now using
\begin{equation}
\begin{aligned}
F_1(t) & = \sum_q e_q \int_{-1}^{1} dx H_q(x,\xi,t) , \ \  \
F_2(t) & = \sum_q e_q \int_{-1}^{1} dx E_q(x,\xi,t) , 
\label{eq:FF-GPD}
\end{aligned}
\end{equation}
and
\begin{equation}
\begin{aligned}
 \rho_1 \left( \left| \mathbf{b_\perp} \right| \right) = \int \frac{d^2 q_\perp}{\left(2 \pi \right)^2} e^{- i \mathbf{q_\perp} \cdot \mathbf{b_\perp}} F_1 \left( q^2_\perp \right), \ \
  \rho_2 \left( \left| \mathbf{b_\perp} \right| \right) = \int \frac{d^2 q_\perp}{\left(2 \pi \right)^2} e^{- i \mathbf{q_\perp} \cdot \mathbf{b_\perp}} F_2 \left( q^2_\perp \right).
  \label{eq:ChDens}
  \end{aligned}
\end{equation}
we get from Eq.\eqref{eq:PartonDist}
\begin{equation}
\rho \left(\mathbf{b_\perp}\right) = \rho_1 \left( \left| \mathbf{b_\perp}\right| \right) -  \frac{1}{2 M_N} \frac{\partial}{\partial y} \rho_2 \left( \left| \mathbf{b_\perp}\right| \right),
\end{equation}
Evaluating the Fourier transform in Eq.\eqref{eq:PartonDist} one  obtains the transverse charge density in impact parameter space~\cite{CarlsonVanderhaeghen1}
\begin{equation}
\rho_T (\mathbf{b_\perp}) = \rho(b_\perp) - \sin(\phi_b - \phi_s) \int_{0}^{\infty} \frac{dq_\perp}{2 \pi} \frac{q^2_{\perp}}{2 M_N} j_1 (b_\perp q_\perp) F_2 (q^2_{\perp}).
\label{eq:Tr-Ch-Den}
\end{equation}
 Tthe charge density in transverse momentum space is obtained by taking the  Fourier transform of Eq.\eqref{eq:PartonDist} (or Eq.\eqref{eq:Tr-Ch-Den})
\begin{equation}
\tilde{\rho}_T (\mathbf{x,q_\perp}) =  H^q (x,-q^2_\perp) + i \frac{q_y}{2 M_N} E^q (x,-q^2_\perp),
\label{eq:Inv-Tr-Ch-Den}
\end{equation}
which implies
\begin{equation}
\tilde{\rho}_T (\mathbf{q_\perp}) =  F_1(q^2_\perp) + i \frac{q_y}{2 M_N} F_2 (q^2_\perp)
\label{eq:Ch-Dens-Mom-Space}
\end{equation}

\section{One and Two Photon Exchange Amplitudes in the Relativistic Eikonal Approximation}
For unpolarized beam of electrons  elastically scattered  from a transversly polarized nucleon traget, the scattering amplitude in the relativistic Eikonal approximation is given by~\cite{Kabat}
\begin{equation}
f^{\uparrow}(\mathbf{q_\perp})= \ - 2 i s \int d^2b_\perp \ e^{-i\mathbf{q_\perp \cdot b_\perp}} \ \left[ e^{i\chi^{\uparrow}(\mathbf{b\perp})}-\ 1\right] ,
\label{eq:Eik1}
\end{equation}
where $s$ is the center of mass energy and $\chi^{\uparrow}\left(\mathbf{{b}_\perp}\right)$ is the Coulomb/Eikonal phase associated with a nucleon target of spin transverse to the incident beam direction. The  momentum transfer is assumed to be purely transverse, i.e.\ $\vec{q_\perp} = \vec{q} - q_z \hat{z}$.
The Coulomb/Eikonal phase is given by 
\begin{equation}
\chi^{\uparrow}(\mathbf{{b}_\perp})=\frac{- 4 \pi \alpha}{2 s } \ \int_{-\infty}^{\infty} dz \ A^{(0)\uparrow}(\mathbf{b_\perp},z) = \frac{- 4 \pi \alpha}{2 s } \ A^{(0)\uparrow}(\mathbf{b_\perp}),
\label{eq:CoulPha}
\end{equation}
here $A^{(0)\uparrow}$ is the electromagnetic transverse potential produced by a transversely polarized nucleon of spin up, and $\alpha =1/137 $ is the electromagnetic coupling constant. Therefore, the Eikonal scattering amplitude becomes after expanding Eq.\eqref{eq:Eik1}
\begin{multline}
f^{\uparrow}(\mathbf{q_\perp}) = \ - 2 i s \int d^2b_\perp \  e^{-i \mathbf{q_\perp \cdot b_\perp}} \ \left[ e^{ \frac{- i 4 \pi \alpha}{2 s } A^{(0) \uparrow}(\mathbf{b_\perp}) }  - \ 1 \right] = \\
- 4 \pi \alpha \int d^2b_\perp  e^{-i\mathbf{q_\perp \cdot b_\perp}}  A^{(0) \uparrow}(\mathbf{b_\perp})  + \frac{i 8 \pi \alpha^2}{s} \int d^2b_\perp  e^{-i\mathbf{q_\perp \cdot b_\perp}}  \left[A^{(0) \uparrow}(\mathbf{b_\perp}) \right]^2 + \cdots.
\label{eq:1-2-pho-exch}
\end{multline}
The above expansion is represented by the following diagram

\begin{figure}[H]
 \begin{center}
 \includegraphics[width=0.6\textwidth]{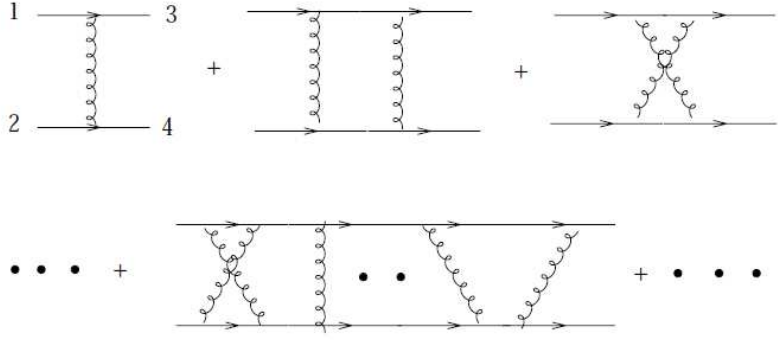} 

\caption{The sum of the diagrams resulting from Eikonal amplitude expansion . Figure from Ref~\cite{DVCS-Ads}.}
 \end{center}
 \label{fig:EikExp}
\end{figure}

\noindent Thus, the first term in the Eikonal approximation (Born approximation) reads
\begin{equation}
f^{\uparrow}_{1\gamma}(\mathbf{q_\perp}) =- 4 \pi \alpha \int d^2b_\perp  e^{-i\mathbf{q_\perp \cdot b_\perp}}  A^{(0) \uparrow}(\mathbf{b_\perp}),
\label{eq:EikAmp11}
\end{equation}
utilizing the Fourier transform of \  $\nabla^2  A^{(0) \uparrow}(\mathbf{b_\perp})     = - \ \rho(\mathbf{b_\perp})$ we get
\begin{equation}
\begin{split}
f^{\uparrow}_{1\gamma}(\mathbf{q_\perp}) = - 4 \pi \alpha \  \tilde{ A}^{(0) \uparrow}(\mathbf{q_\perp}) =- 4 \pi \alpha \ \frac{\tilde{\rho}(\mathbf{q_\perp})}{q_\perp^2}  ,   
\end{split}
\end{equation}
using Eq.\eqref{eq:Ch-Dens-Mom-Space}, the one photon exchange amplitude becomes
\begin{equation}
f^{\uparrow}_{1\gamma}(\mathbf{q_\perp}) = A \left[F_1(q_\perp^2) + \frac{iq_\perp \sin(\phi_{q_\perp} - \  \phi_s)}{2M_N}\ F_2(q_\perp^2)\right]  ,
\label{eq:1PhotAmp}
\end{equation}
where $ A= - \frac{ 4 \pi \alpha }{ q_\perp^2}$
and  $M_N$ is the nucleon mass. Clearly,  the  ($1\gamma$) exchange amplitude depends on the azimuthal angle however, the corresponding cross section is azimuthally symmetric.
\vspace{3mm}
\newline
Next we consider the two photon exchange amplitude, which from Eq.\eqref{eq:1-2-pho-exch} reads
\begin{equation}
f^{\uparrow}_{2\gamma}(\mathbf{q_\perp})= i  \frac{8 \pi \alpha^2}{s} \int d^2b_\perp \ e^{-i\mathbf{q_\perp \cdot b_\perp}} \  \left( {A^{(0) \uparrow}}(\mathbf{b_\perp})  \right)^2 ,
\end{equation} 
from the above formula for $f^{\uparrow}_{2\gamma}(\mathbf{q_\perp})$, we also see that while the amplitude depends on the azimuthal angle, the corresponding cross section  gives no asymmetry (similar to $f^{\uparrow}_{1\gamma}(\mathbf{q_\perp})$ ) in the scattering cross section.
In order to evaluate $f^{\uparrow}_{2\gamma}(\mathbf{q_\perp})$, we rewrite it in the form   
\begin{equation}
f^{\uparrow}_{2\gamma}(\mathbf{q_\perp})= iB \int \frac{d^2b_\perp}{(2 \pi)^2} \ e^{-i\mathbf{q_\perp \cdot b_\perp}} \  A^{(0) \uparrow}(\mathbf{b_\perp})\  A^{(0) \uparrow}(\mathbf{b_\perp}) ,
\end{equation}
where $B= \frac{8 \pi \alpha^2 }{s}$. The above integral represents the Fourier transform of the product of two functions and can  be rewritten using the convolution theorem of Fourier transforms as 
\begin{equation}
\begin{split}
f^{\uparrow}_{2\gamma}(\mathbf{q_\perp})= iB \int \frac{d^2q'_\perp}{(2\pi)^2} \ \tilde{ A}^{(0) \uparrow}(\mathbf{q'_\perp})\ \tilde{ A}^{(0) \uparrow}(\mathbf{q_\perp - \ q'_\perp}) \\
 = iB \int \frac{d^2q'_\perp}{(2\pi)^2} \ \frac{\tilde{\rho}(\mathbf{q'_\perp})}{{\mathbf{q'_\perp}^2}}\ \frac{\tilde{\rho}(\mathbf{q_\perp - \ q'_\perp})}{|\mathbf{q_\perp - \ q'_\perp}|^2},
\end{split}
\end{equation}
using  Eq.\eqref{eq:Ch-Dens-Mom-Space}, $f^{\uparrow}_{2\gamma}(\mathbf{q_\perp})$ becomes 
\begin{equation}
\begin{split}
f^{\uparrow}_{2\gamma}(\mathbf{q_\perp}) = iB \int \frac{d^2q'_\perp}{(2\pi)^2} \frac{1}{\mathbf{|q'_\perp|}^2}\bigg[ F_1(\mathbf{|q'_\perp|}^2) + \frac{iq'_y}{2M_N} F_2(\mathbf{|q'_\perp|}^2)\bigg]   
 \frac{1}{\mathbf{|q_\perp - \ q'_\perp|}^2} \\  \bigg[ F_1(\mathbf{|q_\perp - \ q'_\perp|}^2) +  \frac{i(q_y-q'_y)}{2M_N} F_2(\mathbf{|q_\perp - \ q'_\perp|}^2)\bigg],
 \label{eq:2PhotAmp}
\end{split}
\end{equation}
clearly the two photon exchange contribution appears due to the  convolution between different combinations of  Dirac and Pauli form factors, in contrast to the conventional  $4$-D case where extra form  factors are introduced to define the two photon exchange amplitude. Now from the above equation, we see that $f^{\uparrow}_{2\gamma}(\mathbf{q_\perp}) $ contains the following integrals 
\begin{equation}
I_1 =  \int \frac{d^2q'_\perp}{(2\pi)^2} \frac{1}{\mathbf{|q'_\perp|}^2\mathbf{|q_\perp - \ q'_\perp|}^2} F_1(\mathbf{|q'_\perp|}^2)F_1(\mathbf{|q_\perp - q'_\perp|}^2) ,
\label{eq:I1}
\end{equation}

\begin{equation}
I_2  = \frac{1}{2M_N} \int \frac{d^2q'_\perp}{(2\pi)^2} \frac{(q_y-q'_y)}{\mathbf{|q'_\perp|}^2\mathbf{|q_\perp - \ q'_\perp|}^2} F_1(\mathbf{|q'_\perp|}^2)F_2(\mathbf{|q_\perp - q'_\perp|}^2) ,
\label{eq:I2}
\end{equation}

\begin{equation}
I_3 = \frac{1}{2M_N} \int \frac{d^2q'_\perp}{(2\pi)^2} \frac{q'_y}{\mathbf{|q'_\perp|}^2\mathbf{|q_\perp - \ q'_\perp|}^2} F_1(\mathbf{|q_\perp - q'_\perp|}^2)F_2(\mathbf{|q'_\perp|}^2) ,
\label{eq:I3}
\end{equation}

\begin{equation}
I_4  = \frac{1}{4M^2_N} \int \frac{d^2q'_\perp}{(2\pi)^2} \frac{q'_y(q_y-q'_y)}{\mathbf{|q'_\perp|}^2\mathbf{|q_\perp - \ q'_\perp|}^2} F_2(\mathbf{|q'_\perp|}^2)F_2(\mathbf{|q_\perp - q'_\perp|}^2),
\label{eq:I4}
\end{equation}
where
\begin{equation}
\begin{aligned}
f^{\uparrow}_{2\gamma}(\mathbf{q_\perp}) & = iB \left[ \ I_1 \ + \ i \ I_2 \ + \ i \ I_3 \ - \  I_4 \  \right].
\label{eq:2PhotAmp3}
\end{aligned}
\end{equation}
Performing the  change of variables $\mathbf{q''_\perp=q_\perp-q'_\perp}$ in $I_2$ , one gets
\begin{equation}
I_2= \frac{1}{2M_N} \int \frac{d^2q''_\perp}{(2\pi)^2} \frac{q''_y}{\mathbf{|q''_\perp - \ q_\perp|}^2 \mathbf{|q''_\perp|}^2} F_1(\mathbf{|q''_\perp - q_\perp|}^2)F_2(\mathbf{|q''_\perp|}^2),
\end{equation}
therefore, $I_2 = I_3$, and $f^{\uparrow}_{2\gamma}$ becomes (noting that $q_y = q_\perp \sin(\phi_{q_{\perp}} - \ \phi_s)$)
\begin{equation}
\begin{aligned}
f^{\uparrow}_{2\gamma}(\mathbf{q_\perp}) & = iB \left[ I_1 \ + \ 2   i \  I_3 \ - \  I_4 \right] \\
& = iB \left[ \ I_1 \  + \  I_{41} \ -  \ q_\perp \sin(\phi_{q_{\perp}}  -  \ \phi_s) \  I_{42}   \ 
  + \ 2  i  \   I_3 \ \right]  \\
& = iB \left[ \ I_1 \  + \  I_{41} \ -  \ q_\perp \sin(\phi_{q_{\perp}}  -  \ \phi_s) \  I_{42}  \ \right]  -  2  B \ I_3,
\label{eq:2PhotAmp2}
\end{aligned}
\end{equation}
where 
\begin{equation}
\begin{aligned}
I_{41} & = \frac{1}{4M^2_N} \int \frac{d^2q'_\perp}{(2\pi)^2} \frac{q'^{2}_{y}}{\mathbf{|q'_\perp|}^2\mathbf{|q_\perp - \ q'_\perp|}^2} F_2(\mathbf{|q'_\perp|}^2)F_2(\mathbf{|q_\perp - q'_\perp|}^2), \\
I_{42} &= \frac{1}{4M^2_N} \int \frac{d^2q'_\perp}{(2\pi)^2} \frac{q'_y}{\mathbf{|q'_\perp|}^2\mathbf{|q_\perp - \ q'_\perp|}^2} F_2(\mathbf{|q'_\perp|}^2)F_2(\mathbf{|q_\perp - q'_\perp|}^2).
\label{eq:I4-12}
\end{aligned}
\end{equation}
Note that  $I_{2}$,  $I_{41}$ and $I_{42}$ are free from  $IR$ divergences (considering polar coordinates and  the symmetry between the poles at $q'_\perp=0$ and $q'_\perp = q_\perp$), while $I_1$ is $IR$ divergent. To extract the $IR$ divergence in $I_1$ we add a photon mass to the divergent part and use dimensional regularization, as illustrated in the next section.

\section{Isolating the $IR$ Divergence in the Two Photon Exchange Amplitude Using Dimensional Regularization}

The integrals $I_{41}$ and $I_{42}$ in Eq.\eqref{eq:I4-12} are free from $IR$ divergences and can be evaluated numerically, while $I_1$ in  Eq.\eqref{eq:I1}  contains  $IR$ divergence that one needs to deal with. In the rest of this section (and appendix A), we use dipole parametrization for $F_1(q^2_\perp)$ and $F_2(q^2_\perp)$ to show that the behavior of the $IR$ divergence in impact paraemter space is similar to that in conventional $4$-D space~\cite{AfanaseveTPE1,Arrington-2PE,Blunden2-2PE,Peskin}, where it was found that the $IR$ divergence cancels with the Bremsstrahlung contribution to the two photon amplitude (The details of the Bremsstrahlung calculations and the proof that it cancels with the $IR$ divergence in the $2 \gamma$ amplitude are  found in the above references). Thus it is tempting to start by studying the behavior of the $IR$ divergence of the $2 \gamma$ amplitude in ampact parameter space and compare the result with the $4$-D case, to do this we  first use the following dipole parametrization for the form factors~\cite{DirPauFF}, (Sachs form factors can be used as well and the same analysis used here can be employed)
\begin{equation}
\begin{aligned}
F_1(q^2_{\perp}) = \frac{F_1(0)}{\left(1+q^2_{\perp}\big/ M^2_d \right)^2}, \ \ \
F_2(q^2_{\perp}) = \frac{F_2(0)}{\left(1+q^2_{\perp}\big/ M^2_d \right)^2},
\end{aligned}
\end{equation}
where $F_1(0)=1$, $F_2(0) = \kappa_p =1.79$, $M^2_d = 0.71 \ GeV^2$.
Introducing  a virtual photon mass $m'$ to regularize the $IR$ divergence at $q'_\perp =0$ and  $q'_\perp = q_\perp$ and using dimensional regularization with $D=2-\epsilon$, $I_1$ becomes
\begin{multline}
I_1  =   \int \frac{d^Dq'_\perp}{(2\pi)^D} \frac{1}{\left( \mathbf{|q'_\perp|}^2 +m'^2 \right) \left( \mathbf{|q'_\perp - \ q_\perp|}^2 +m'^2 \right)} \times  \frac{M^8_d}{\left( \mathbf{|q'_\perp|}^2 +M^2_d \right)^2 \left( \mathbf{|q'_\perp - \ q_\perp|}^2 +M^2_d \right)^2}.
\label{eq:I1_dipole}
\end{multline}
Using partial fractions, the above integral can be decomposed into $(9)$ integrals (the same result can be obtained using the package FeynCalc~\cite{FeynCalc}) which can be  evaluated using dimensional regularization (appendix A contains detailed calculations), where all the $IR$ divergences cancel except a logarithmic term that appears due to the following integral ($I_{15}$ in appendix A ) 
\begin{equation}
I_{1_{IR}}=\frac{M^8_d}{(m'^2-M^2_d)^4} \int \frac{d^D \mathbf{ q'_\perp}}{\left(2 \pi \right)^D} \frac{1}{\left(m'^2 + \mathbf{ q'^2_{\perp}} \right) \left[m'^2 + (\mathbf{q'_\perp} - \mathbf{q_\perp)}^2 \right]},
\end{equation}
using Feynman parametrization  we get
\begin{equation}
\begin{aligned}
I_{1_{IR}} =  & \frac{M^8_d \Gamma(1)}{(m'^2-M^2_d)^3 (2\pi)^{1-\frac{\epsilon}{2}} \Gamma(1) \Gamma(1)} \int_0^1 \frac{1}{\left[m'^2 + q^2_\perp x (1-x) \right]^{1+ \frac{\epsilon}{2}}} dx \\
 = & \frac{ 2 M^8_d}{2 \pi (m'^2-M^2_d)^3 q_\perp \sqrt{4 m'^2+q^2_\perp}} \ln \left(\frac{q_\perp \left(\sqrt{4 m'^2+q^2_\perp}+q_\perp \right)+2 m'^2}{2 m'^2}\right)  \\ & \approx \frac{-  M^2_d}{ \pi q^2_\perp} \ln \left( \frac{q^2_\perp}{m'^2} \right).
 \label{eq:leading-order2}
\end{aligned}
\end{equation}
Similar decomposition can be done for $I_{41}$ and $I_{42}$ in Eq.\eqref{eq:I4-12} from which one can show, using dimensional regularization, that all the $IR$ divergences cancel each other, however, $I_{41}$ and  $I_{42}$ can be evaluated numerically using polar coordinates. On the other hand, the logarithmic  result of the $IR$ divergence of the $2 \gamma$ amplitude in impact parameter space is similar to that in the $4-D$ case~\cite{Arrington-2PE,Blunden2-2PE}. In appendix A, we show using dimensional regularization the full evaluation of $I_1$ using the dipole form factors where it is noted that  all the $IR$ divergences  cancel except a logarithmic term that   appears in the following integral (see appendix A for the details)
\begin{equation}
\begin{aligned}
I_{15} & = \frac{1}{(m'^2-M^2_d)^4} \int \frac{d^D \mathbf{ q'_\perp}}{\left(2 \pi \right)^D} \frac{1}{\left(m'^2 + \mathbf{ q'^2_{\perp}} \right) \left[m'^2 + (\mathbf{q'_\perp} - \mathbf{q_\perp)}^2 \right]} \\
  & = \frac{\Gamma(1)}{(m'^2-M^2_d)^3 (2\pi)^{1-\frac{\epsilon}{2}} \Gamma(1) \Gamma(1)} \int_0^1 \frac{1}{\left[m'^2 + q^2_\perp x (1-x) \right]^{1+ \frac{\epsilon}{2}}} dx\\
  & = \frac{1}{2\pi (m'^2-M^2_d)^3 }  \frac{2 \ln \left(\frac{q \left(\sqrt{4 m'^2+q^2}+q\right)+2 m'^2}{2 m'^2}\right)}{q \sqrt{4 m'^2+q^2}},
\end{aligned}
\end{equation}
this result reduces to the result obtained in Eq.\eqref{eq:leading-order2} using the leading order of the integral. Again we emphasis that the full analysis shown in appendix $A$ can be applied to Sachs form factors, however the number of integrals resulting from using dimensional regularization will be large. Therefore it is useful to try to evaluate the $2 \gamma$ exchange  integrals numerically, which also allow us to use other parametrizations of the form factors, such that generalized parton distributions parametrization, $GPDs$. The only $IR$ divergence that we need to take into account is that appears in $I_1$ since in the other integrals the divergences canceled when using polar coordinates and the symmetry of the propagators. In the following section we show a possible way deal with such divergence.

\section{Evaluation of the Two Photon Exchange Amplitude for Arbitrary Parametrization of the Form Factors}
The integrals in Eqs.\eqref{eq:2PhotAmp2} can be evaluated numerically  for any parametrization of the form factors. For $I_1$, the $IR$ divergence can be extracted by expanding the propagator $ \frac{1}{\mathbf{|q'_\perp -  \ q_\perp|^2}}$  in Eq.\eqref{eq:I1} (see appendix B for  the details)
\begin{multline}
I_1 = \frac{1}{q^2_\perp}  \int \frac{d\phi' dq'_\perp}{(2\pi)^2} \left[ \   \frac{2}{q_\perp} \cos(\phi - \phi') \ + \   \frac{q'_\perp}{q^2_\perp} \left( \cos^2(\phi - \phi') \  + \ 3 \cos(\phi - \phi') \  - 1 \ \right) \ + \cdots   \right] \times  \\ F_1(\mathbf{|q'_\perp|}^2)F_1(\mathbf{|q'_\perp - q_\perp|}^2)
\label{eq:Exp_I1}
\end{multline}
The above expansion follows from the addition theorem of spherical harmonics for $q'_\perp < q_\perp $ for which the  charge density is represented by the transverse charge density of the nucleon. In the numerical calculations, the size of the nucleon in the transverse (or impact parameter)  plane was taken from references~\cite{CarlsonVanderhaeghen1,MillerRev} and we used the leading order of the finite part of the above expansion after subtracting the $IR$ divergence that leads to the logarithmic divergence which at the end cancels with the Bremsstrhalung contribution as shown in the previous section.

\vspace{3mm}

\begin{figure}[H]
 \begin{center}
  \includegraphics[width=0.45\textwidth]{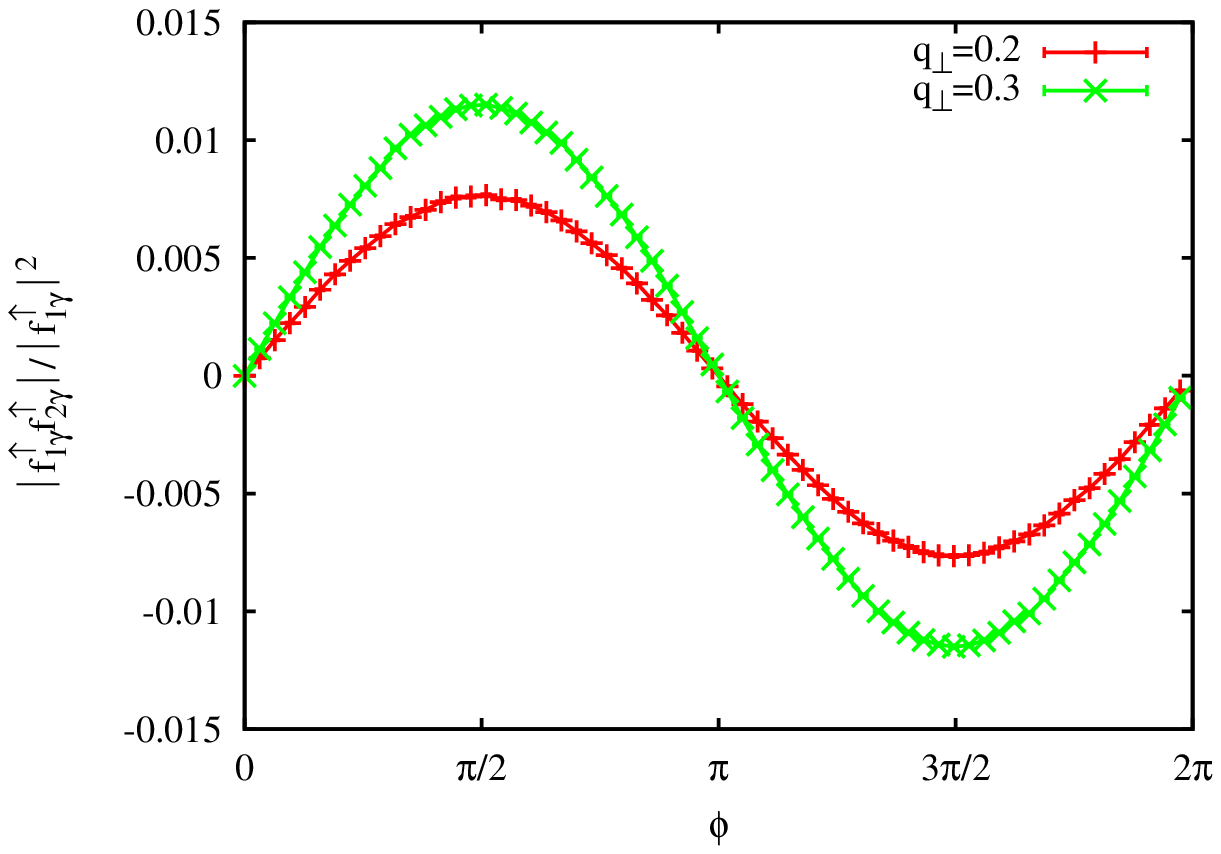}
   \includegraphics[width=0.45\textwidth]{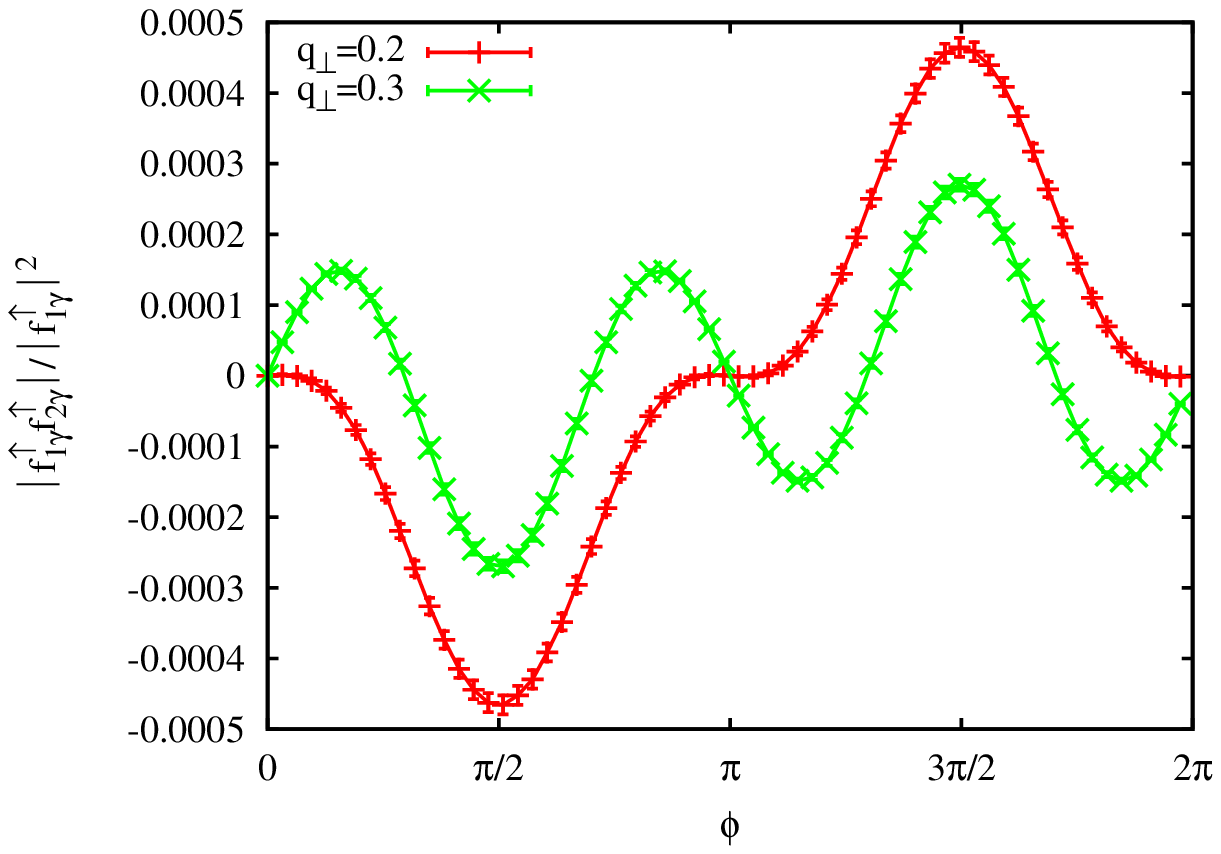}
 \end{center}
\caption{Interference of $1 \gamma$ and $2\gamma$ amplitudes normalized to Born  cross section for proton (left) and neutron (right)}
 \label{fig:Interference-1-2-P-N}
\end{figure}

\begin{figure}[H]
 \begin{center}
  \includegraphics[width=0.45\textwidth]{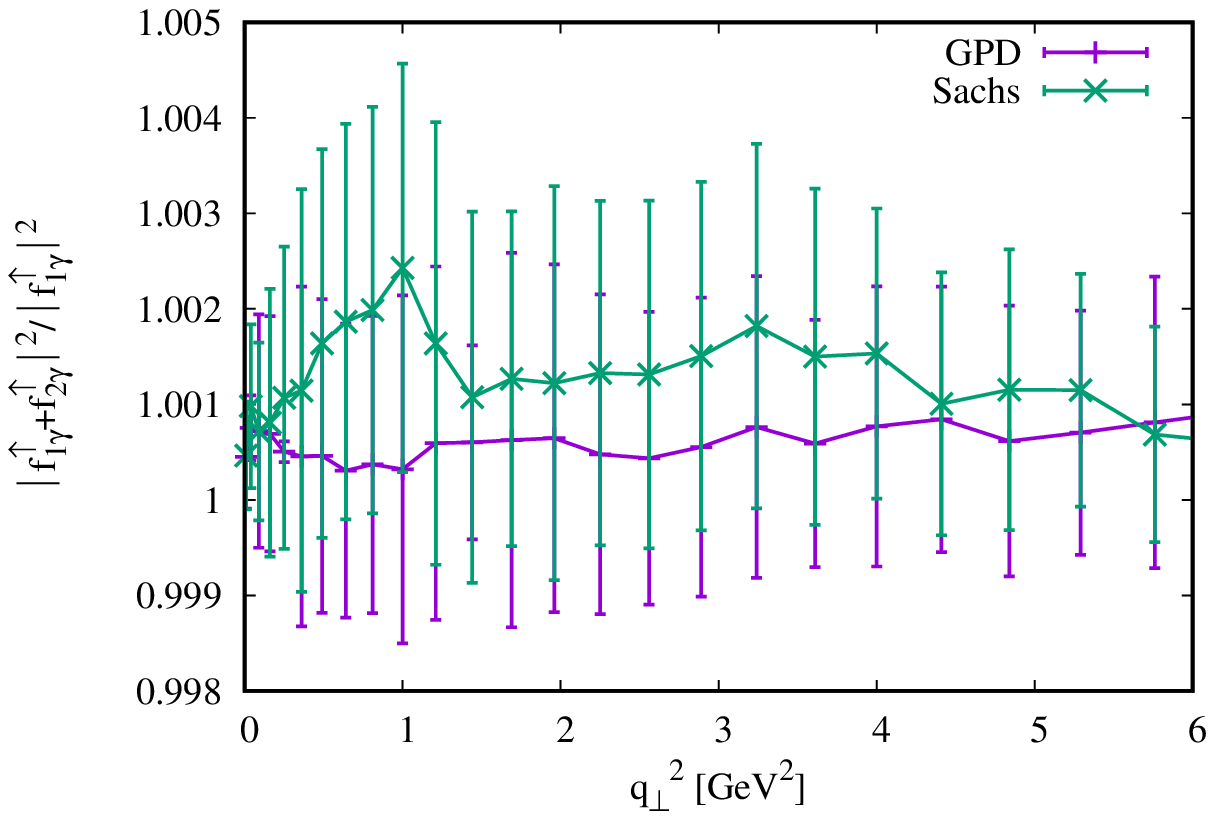}
   \includegraphics[width=0.45\textwidth]{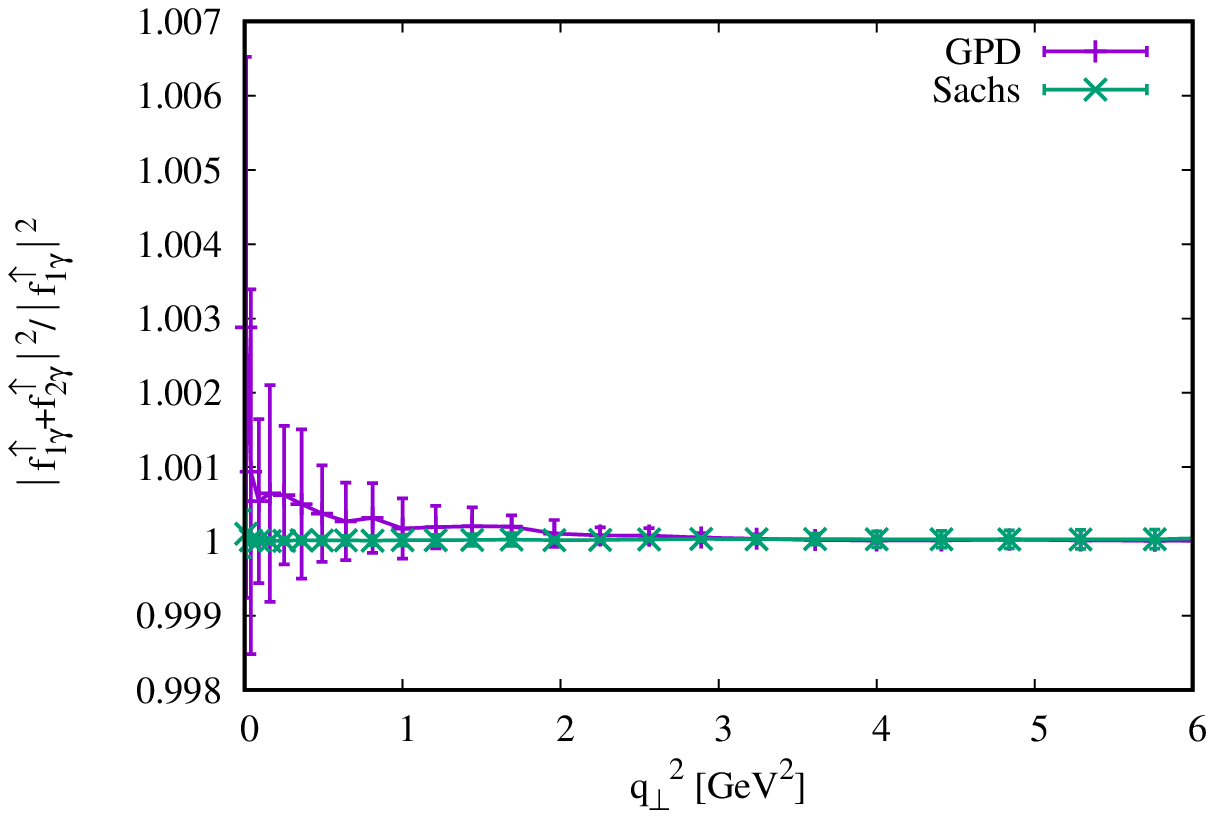}
 \end{center}
\caption{ Ratio of the Born + $2\gamma$ contribution to elastic cross section for proton (left) and neutron (right) relative to Born term, using GPD and Sachs parametrization of the form factors}
 \label{fig:Ratio-P-N}
\end{figure}

\begin{figure}[H]
 \begin{center}
  \includegraphics[width=0.45\textwidth]{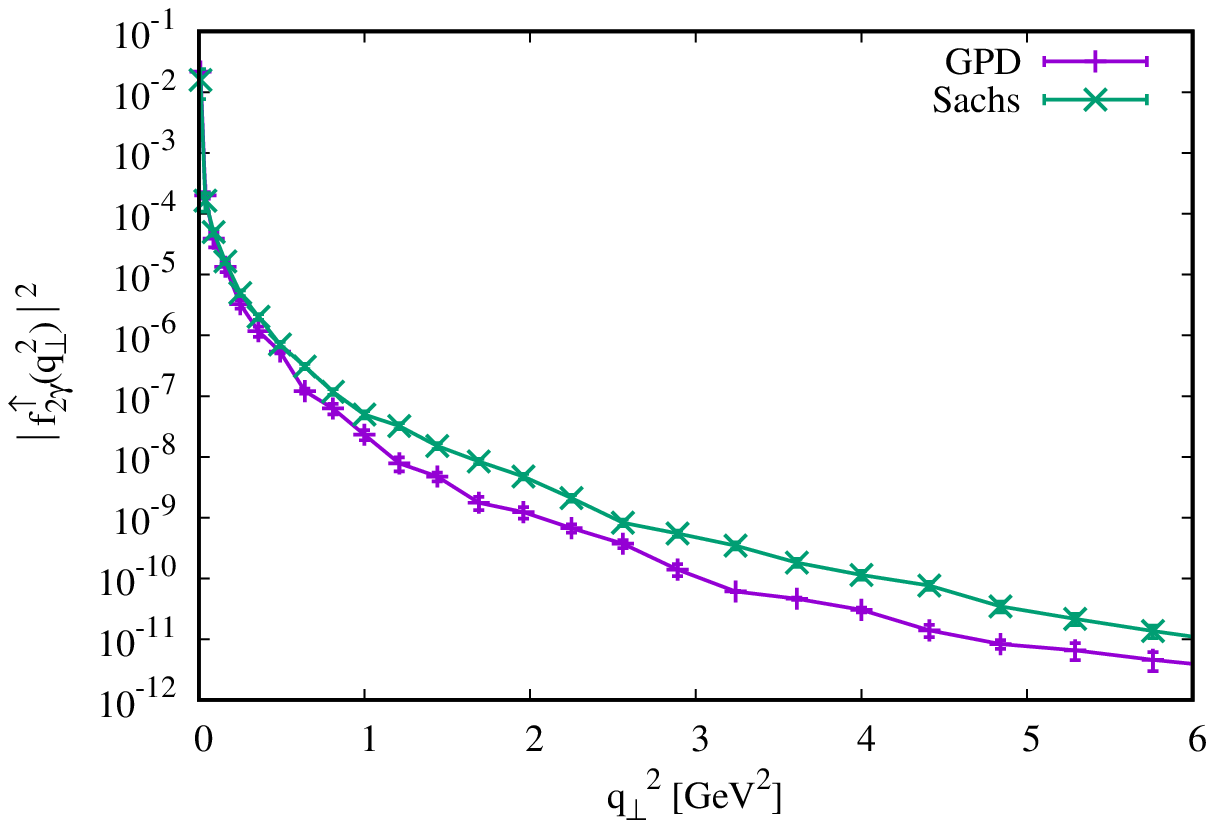}
   \includegraphics[width=0.45\textwidth]{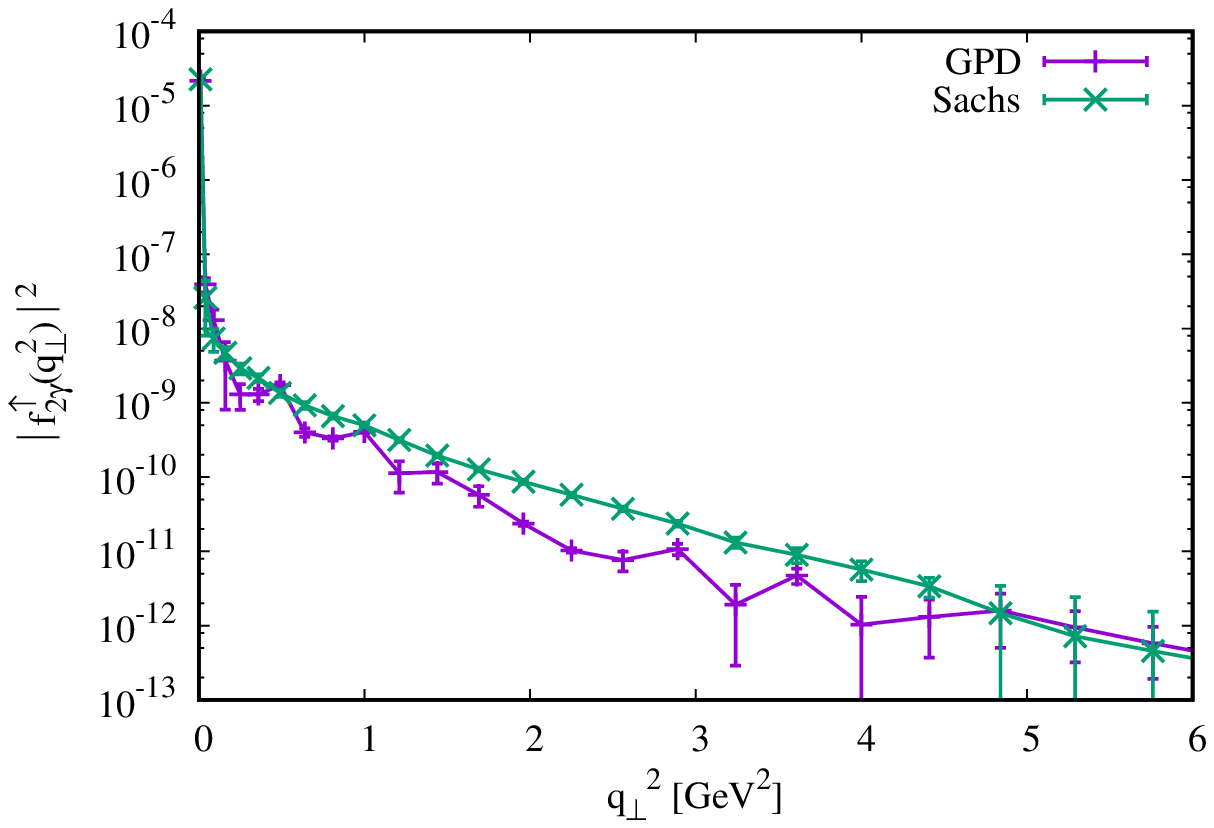}
 \end{center}
\caption{ $2\gamma$ contribution  to elastic cross section (in $GeV^{-2}$) for proton (left) and neutron (right) using GPD and Sachs parametrization of the form factors}
 \label{fig:2-Photon-Amp-P-N}
\end{figure}

As shown in section $III$, the one and two photon amplitudes are azimuthally symmetric and therefore we do not expect any asymmetry from the corresponding cross sections. However, the interference term of the $1 \gamma$ and $2 \gamma$ amplitudes  is azimuthally asymmetric, and $SSA$ appears due to this term.  Figure \ref{fig:Interference-1-2-P-N} shows the  numerical calculations of the azimuthal distribution of the interference between the $1 \gamma$ and $2 \gamma$ amplitudes normalized to the $1 \gamma$ (Born) cross section for proton and neutron at two different momentum transfers. In Figure \ref{fig:Ratio-P-N} the ratio of the total cross section to the Born term is shown which is consisten with the 4-D case \cite{Blunden-2PE}, the $2\gamma$ contribution (for proton and neutron) to the elastic scattering with nucleon intermediate state is shown in Figure \ref{fig:2-Photon-Amp-P-N}. Since single spin asymmetry is proportional to the interference of the $1 \gamma$ and $2 \gamma$ amplitudes \cite{Blunden-2PE, Carlson-Vander-TPE},  the amplitudes  in Figure \ref{fig:Interference-1-2-P-N} are a measure to this asymmetry. On the other hand, these plots show opposite signs for proton and neutron, which is an indication of the sign of azimuthal $SSA$ for unpolarized electrons scattered from transversely  polarized nucleon and is consistent with the proton results\cite{AfanaseveTPE1, Blunden-2PE} and neutron results  (using transversely polarized  $^3He$) in recent Jlap measurements for neutron \cite{Zhang1}. A more detailed study of the transverse target  azimuthal $SSA$ using the transverse electromagnetic potential associated with a transversely polarized nucleons  is currently under preparation. The parametrization of the Sachs $G_E$ and $G_M$ form factors (appendix $C_{1}$ ) were taken from Ref\cite{Alberico} and the $GPD$ parametrizations (appendix $C_{2}$) used were taken from Ref\cite{GPD2}.

\section{Conclusion}
In this paper, we calculate the one and two photon exchange Eikonal amplitudes in impact parameter space for elastic $e-N^\uparrow$ scattering. The results show that the two photon exchange amplitude appears due to the convolution between different combinations of Dirac and Pauli form factors. On the other hand, while the amplitudes associated with the one and two photon exchanges are azimuthally asymmetric in the transverse plane, the corresponding cross sections are not. However, the interference term between the $1 \gamma$ and $2 \gamma$ amplitudes is azimuthally asymmetric, which is consistent with elastic and deep inelastic scattering for the $4$-D  case, and an indication of the existence of azimuthal single spin asymmetry in elastic scattering for both proton and neutron which  can be attributed to the fact that the nucleon charge density is transversely (azimuthally) distorted in the transverse plane for transversely polarized nucleons. It is also noted that the two photon amplitude in impact parameter space contains an $IR$ divergence that has the same logarithmic behavior as in the $4$-D case, which cancels with the Bremsstrahlung contribution to the cross section due to real photon emission from the target and the beam. 

\vspace{3mm}
As a future work and a direct consequence of the results of this paper, is the calculation of the target azimuthal single spin asymmetry in elastic scattering for transversely   polarized protons and neutrons utilizing a recent calculation of the corresponding transverse electromagnetic potential.  Another possible consequence of this work is to extend the calculations of  the $2 \gamma $ exchange amplitude in transverse plane to  different processes in deep inelastic scattering, which should lead to the calculation of $SSA$ for such processes.

\vspace{5mm}

\appendixtitleon
\appendixtitletocon

\begin{appendices}

\section{Full Dimensional Regularization Evaluation of $I_1$ for the Dipole Parametrization in Eq.\eqref{eq:I1_dipole}}
We start by decomposing the integrand of $I_1$ given in Eq.\eqref{eq:I1_dipole} using partial fractions, this will allow us to easily  use dimensional regularization to evaluate $I_1$, noting that 
\begin{equation}
\frac{1}{(x^2+a)(x^2+b)^2} = \frac{1}{(a-b)(b+x^2)^2}+\frac{1}{(a-b)^2(a+x^2)}-\frac{1}{(a-b)^2(b+x^2)},
\end{equation}
we have
\begin{multline}
\frac{1}{(\mathbf{q_\perp^2} + m'^2)(\mathbf{q_\perp^2} + M^2_d)^2} =  \frac{1}{(m'^2-M^2_d)(M^2_d+ \mathbf{q_\perp^2})^2} +  \frac{1}{(m'^2-M^2_d)^2 (m'^2+ \mathbf{q_\perp^2})} - \\  \frac{1}{(m'^2-M^2_d)^2 (M^2_d+ \mathbf{q_\perp^2})} ,
\end{multline} 
and
\begin{multline}
\frac{1}{ \left[ \mathbf{(q'_{\perp} - q_\perp})^2 + m'^2 \right] \left[ \mathbf{(q'_{\perp} -q_\perp})^2 + M^2_d \right]^2} = \\  \frac{1}{(m'^2-M^2_d)(M^2_d+ \left[ \mathbf{(q'_{\perp} -q_\perp})^2 + M^2_d \right]^2} +  \frac{1}{(m'^2-M^2_d)^2 (m'^2+ \left[ \mathbf{(q'_{\perp}-q_\perp})^2 + M^2_d \right]^2}   \\ - \frac{1}{(m'^2-M^2_d)^2 (M^2_d+ \left[ \mathbf{(q'_{\perp}-q_\perp})^2 + M^2_d \right]^2} 
\end{multline}
Therefore $I_1$,  (Eq.\eqref{eq:I1_dipole}) consists of the following integrals
\begin{equation}
I_{11}=\frac{1}{(m'^2-M^2_d)^2} \int \frac{d^D \mathbf{ q'_\perp}}{\left(2 \pi \right)^D} \frac{1}{\left(M^2_d + \mathbf{ q'^2_{\perp}} \right)^2 \left[M^2_d + (\mathbf{q'_\perp} - \mathbf{q_\perp)}^2 \right]^2},
\end{equation}
\begin{equation}
I_{12}=\frac{1}{(m'^2-M^2_d)^3} \int \frac{d^D \mathbf{ q'_\perp}}{\left(2 \pi \right)^D} \frac{1}{\left(M^2_d + \mathbf{ q'^2_{\perp}} \right)^2 \left[m'^2 + (\mathbf{q'_\perp} - \mathbf{q_\perp)}^2 \right]},
\end{equation}
\begin{equation}
I_{13}=-\frac{1}{(m'^2-M^2)^3} \int \frac{d^D \mathbf{ q'_\perp}}{\left(2 \pi \right)^D} \frac{1}{\left(M^2_d + \mathbf{ q'^2_{\perp}} \right)^2 \left[M^2_d + (\mathbf{q'_\perp} - \mathbf{q_\perp)}^2 \right]},
\end{equation}
\begin{equation}
I_{14}=\frac{1}{(m'^2-M^2_d)^3} \int \frac{d^D \mathbf{ q'_\perp}}{\left(2 \pi \right)^D} \frac{1}{\left(m'^2 + \mathbf{ q'^2_{\perp}} \right) \left[M^2_d + (\mathbf{q'_\perp} - \mathbf{q_\perp)}^2 \right]^2},
\end{equation}
\begin{equation}
I_{15}=\frac{1}{(m'^2-M^2_d)^4} \int \frac{d^D \mathbf{ q'_\perp}}{\left(2 \pi \right)^D} \frac{1}{\left(m'^2 + \mathbf{ q'^2_{\perp}} \right) \left[m'^2 + (\mathbf{q'_\perp} - \mathbf{q_\perp)}^2 \right]},
\end{equation}
\begin{equation}
I_{16}=-\frac{1}{(m'^2-M^2_d)^4} \int \frac{d^D \mathbf{ q'_\perp}}{\left(2 \pi \right)^D} \frac{1}{\left(m'^2 + \mathbf{ q'^2_{\perp}} \right) \left[M^2_d + (\mathbf{q'_\perp} - \mathbf{q_\perp)}^2 \right]},
\end{equation}
\begin{equation}
I_{17}=-\frac{1}{(m'^2-M^2_d)^3} \int \frac{d^D \mathbf{ q'_\perp}}{\left(2 \pi \right)^D} \frac{1}{\left(M^2_d + \mathbf{ q'^2_{\perp}} \right) \left[M^2_d + (\mathbf{q'_\perp} - \mathbf{q_\perp)}^2 \right]^2},
\end{equation}
\begin{equation}
I_{18}=-\frac{1}{(m'^2-M^2_d)^4} \int \frac{d^D \mathbf{ q'_\perp}}{\left(2 \pi \right)^D} \frac{1}{\left(M^2_d + \mathbf{ q'^2_{\perp}} \right) \left[m'^2 + (\mathbf{q'_\perp} - \mathbf{q_\perp)}^2 \right]},
\end{equation}
\begin{equation}
I_{19}=\frac{1}{(m'^2-M^2_d)^4} \int \frac{d^D \mathbf{ q'_\perp}}{\left(2 \pi \right)^D} \frac{1}{\left(M^2_d + \mathbf{ q'^2_{\perp}} \right) \left[M^2_d + (\mathbf{q'_\perp} - \mathbf{q_\perp)}^2 \right]}.
\end{equation}
Using  Feynman parametrization, and evaluating the momentum integrals for $D=2-\epsilon$, one gets

\begin{equation}
\begin{aligned}
I_{11} & = \frac{\Gamma(3)}{(m'^2-M^2_d)^2 (2\pi)^{1-\frac{\epsilon}{2}} \Gamma(2) \Gamma(2)} \int_0^1 \frac{(1-x)x}{\left[M^2_d + q^2_\perp x (1-x) \right]^{3+ \frac{\epsilon}{2}}} dx,  \\
I_{12} & = \frac{\Gamma(2)}{(m'^2-M^2_d)^3 (2\pi)^{1-\frac{\epsilon}{2}} \Gamma(2) \Gamma(1)} \int_0^1 \frac{(1-x)}{\left[M^2_d + q^2_\perp x (1-x) +(m'^2-M^2_d)x \right]^{2 + \frac{\epsilon}{2}}} dx,  \\
I_{13} & = -\frac{\Gamma(2)}{(m'^2-M^2_d)^3 (2\pi)^{1-\frac{\epsilon}{2}} \Gamma(2) \Gamma(1)} \int_0^1 \frac{(1-x)}{\left[M^2_d + q^2_\perp x (1-x) \right]^{2+ \frac{\epsilon}{2}}} dx , \\
I_{14} & = \frac{\Gamma(2)}{(m'^2-M^2_d)^3 (2\pi)^{1-\frac{\epsilon}{2}} \Gamma(2) \Gamma(1)} \int_0^1 \frac{(1-x)}{\left[m'^2 + q^2_\perp x (1-x) +(M^2_d-m'^2)x \right]^{2 + \frac{\epsilon}{2}}} dx, \\
I_{15} & = \frac{\Gamma(1)}{(m'^2-M^2_d)^3 (2\pi)^{1-\frac{\epsilon}{2}} \Gamma(1) \Gamma(1)} \int_0^1 \frac{1}{\left[m'^2 + q^2_\perp x (1-x) \right]^{1+ \frac{\epsilon}{2}}} dx,  \\
I_{16} & = \frac{- \Gamma(1)}{(m'^2-M^2_d)^4 (2\pi)^{1-\frac{\epsilon}{2}} \Gamma(2) \Gamma(1)} \int_0^1 \frac{1}{\left[m'^2 + q^2_\perp x (1-x) +(M^2_d-m'^2)x \right]^{1 + \frac{\epsilon}{2}}} dx , \\
I_{17} & = -\frac{\Gamma(2)}{(m'^2-M^2_d)^3 (2\pi)^{1-\frac{\epsilon}{2}} \Gamma(1) \Gamma(2)} \int_0^1 \frac{x}{\left[M^2_d + q^2_\perp x (1-x) \right]^{2+ \frac{\epsilon}{2}}} dx,  \\
I_{18} & = \frac{- \Gamma(1)}{(m'^2-M^2_d)^4 (2\pi)^{1-\frac{\epsilon}{2}} \Gamma(1) \Gamma(1)} \int_0^1 \frac{1}{\left[M^2 + q^2_\perp x (1-x) +(m'^2-M^2_d)x \right]^{1 + \frac{\epsilon}{2}}} dx,  \\
I_{19} & = \frac{\Gamma(1)}{(m'^2-M^2_d)^4 (2\pi)^{1-\frac{\epsilon}{1}} \Gamma(1) \Gamma(1)} \int_0^1 \frac{1}{\left[M^2 + q^2_\perp x (1-x) \right]^{1+ \frac{\epsilon}{2}}} dx.
\end{aligned}
\end{equation}
Using Mathematica, one obtains for the integrals over $x$ (these are the above integrals without the mass factor in the left hand side of each integrals)
\begin{multline}
I_{11x} = \left( q \left(q^2-2 M^2\right) \sqrt{4 M^2+q^2}+2 M^2 \left(M^2+q^2\right) \ln \left(\frac{\sqrt{4 M^2+q^2}+q}{\sqrt{4 M^2+q^2}-q}\right) -  \right. \\  \left. 2 M^2 \left(M^2+q^2\right) \ln \left(\frac{\sqrt{4 M^2+q^2}-q}{\sqrt{4 M^2+q^2}+q}\right) \right) \bigg/ M^2 q^3 \left(4 M^2+q^2\right)^{5/2} ,
\end{multline}

\begin{equation}
\begin{aligned} 
I_{12x}   = &  \left[  \left(m'^2-M^2+q^2\right) \sqrt{m'^4-2 m'^2 \left(M^2-q^2\right)+\left(M^2+q^2\right)^2} \ - \right.   \\ & \left. M^2 \left(-m'^2+M^2+q^2\right) \times \right. \\ & \left. \ln \left(\frac{-m'^2+\sqrt{m'^4-2 m'^2 \left(M^2-q^2\right)+\left(M^2+q^2\right)^2}+M^2-q^2}{m'^2+\sqrt{m'^4-2 m'^2 \left(M^2-q^2\right) +  \left(M^2+q^2\right)^2}-M^2+q^2}\right) - \right.  \\ &  \left.  M^2 \left(-m'^2+M^2+q^2\right) \times \right. \\ &  \left.  \ln \left(\frac{m'^2+\sqrt{m'^4-2 m'^2 \left(M^2-q^2\right)+  \left(M^2+q^2\right)^2}-M^2-q^2}{-m'^2+\sqrt{m'^4-2 m'^2 \left(M^2-q^2\right)+\left(M^2+q^2\right)^2}+M^2+q^2}\right) \right]   \\ & \bigg/   M^2 \left(m'^4-2 m'^2 \left(M^2-q^2\right)+\left(M^2+q^2\right)^2\right)^{3/2} 
\end{aligned}
\end{equation}
\begin{align}
 I_{13x} = &  \frac{4 \tanh ^{-1}\left(\frac{q}{\sqrt{4 M^2+q^2}}\right)}{q \left(4 M^2+q^2\right)^{3/2}}+\frac{1}{4 M^4+M^2 q^2}, &
\end{align}
\begin{equation}
\begin{aligned}
I_{14x} = & \left[  \left(-m'^2+M^2+q^2\right) \sqrt{m'^4-2 m'^2 \left(M^2-q^2\right)+\left(M^2+q^2\right)^2} \right. \\ & \left. -  m'^2 \left(m'^2-M^2+q^2\right) \times \right. \\ &  \left. \ln \left(\frac{-m'^2+\sqrt{m'^4-2 m'^2 \left(M^2-q^2\right)+\left(M^2+q^2\right)^2}+M^2-q^2}{m'^2+\sqrt{m'^4-2 m'^2 \left(M^2-q^2\right)+\left(M^2+q^2\right)^2}- M^2+q^2}\right) \right.  \\ & \left. -m'^2  \left(m'^2-M^2+q^2\right)  \times \right. \\ & \left.  \ln \left(\frac{m'^2+\sqrt{m'^4-2 m'^2 \left(M^2-q^2\right)+\left(M^2+q^2\right)^2}-M^2-q^2}{-m'^2+\sqrt{m'^4-2 m'^2 \left(M^2-q^2\right)+\left(M^2+q^2\right)^2}+M^2+q^2}\right) \right]  \\ & \bigg/  m'^2 \left(m'^4-2 m'^2 \left(M^2-q^2\right)+\left(M^2+q^2\right)^2\right)^{3/2},
\end{aligned}
\end{equation}
\begin{align}
I_{15x}  = &  \frac{2 \ln \left(\frac{q \left(\sqrt{4 m'^2+q^2}+q\right)+2 m'^2}{2 m'^2}\right)}{q \sqrt{4 m'^2+q^2}} , &
\end{align}
\begin{align}
I_{16x}  = &  \frac{\ln \left(\frac{m'^2+\sqrt{m'^4-2 m'^2 \left(M^2-q^2\right)+\left(M^2+q^2\right)^2}+M^2+q^2}{m'^2-\sqrt{m'^4-2 m'^2 \left(M^2-q^2\right)+\left(M^2+q^2\right)^2}+M^2+q^2}\right)}{\sqrt{m'^4-2 m'^2 \left(M^2-q^2\right)+\left(M^2+q^2\right)^2}} , &
\end{align}
\begin{align}
I_{17x} = &  \frac{4 \tanh ^{-1}\left(\frac{q}{\sqrt{4 M^2+q^2}}\right)}{q \left(4 M^2+q^2\right)^{3/2}}+\frac{1}{4 M^4+M^2 q^2} , &
\end{align}
\begin{align}
I_{18x} = & \frac{\ln \left(\frac{m'^2+\sqrt{m'^4-2 m'^2 \left(M^2-q^2\right)+\left(M^2+q^2\right)^2}+M^2+q^2}{m'^2-\sqrt{m'^4-2 m'^2 \left(M^2-q^2\right)+\left(M^2+q^2\right)^2}+M^2+q^2}\right)}{\sqrt{m'^4-2 m'^2 \left(M^2-q^2\right)+\left(M^2+q^2\right)^2}} , &
\end{align}
\begin{align}
I_{19x} = & \frac{2 \ln \left(\frac{q \left(\sqrt{4 M^2+q^2}+q\right)+2 M^2}{2 M^2}\right)}{q \sqrt{4 M^2+q^2}}. &
\end{align}

\section{Expansion of \ $\frac{1}{|q'_\perp - q_\perp|}$  in Impact Parameter Space}

The impact parameter space expansion of  $\frac{1}{|q'_\perp - q_\perp|}$ can be obtained using
\begin{equation}
\frac{1}{|\mathbf{q_\perp} - \mathbf{q'_\perp} |} = \sum_{l \geq 0}  \sum_{m=-l}^{l} \frac{4 \pi}{2 l + 1} \frac{q^l_{\perp <}}{q^{' l+1}_{\perp >}} \tilde{Y}_{lm} (\frac{\pi}{2},\phi') Y_{lm} (\frac{\pi}{2},\phi),
 \end{equation}
 where
 \begin{equation}
 \begin{aligned}
 P_l(\cos \gamma) \ = & \  \frac{4 \pi}{2 l + 1}  \sum_{m=-l}^{l} \tilde{Y}_{lm} (\theta',\phi') Y_{lm} (\theta,\phi) \\
 P_0(x) = &  \ 1, P_1(x) = x, P_2(x) = \frac{1}{2} (3 x^2 -1), P_3(x) = \frac{1}{2} (5 x^3 - 3 x),  \\
 \cos(\gamma) = & \sin (\theta) \sin(\theta') \cos(\phi - \phi') + \cos(\theta) \cos(\theta'),
 \end{aligned}
 \end{equation}
 for $\theta= \frac{\pi}{2}$ we get
 \begin{equation}
 \cos(\gamma) =  \cos(\phi - \phi') 
  \end{equation}
 Thus the expansion of $\frac{1}{|q'_\perp - q_\perp|}$ becomes
 \begin{equation}
 \frac{1}{|q'_\perp - q_\perp|} = \frac{1}{q_\perp} \left[ 1 + \frac{q'_\perp}{q_\perp} \cos(\phi - \phi')  + \left(  \frac{q'_\perp}{q_\perp} \right)^2 \left(  \frac{3 \cos^2(\phi - \phi')  }{2} - \frac{1}{2}\right) + \cdots  \right]
 \end{equation}

\section{Form Factors Parametrizations}

\subsection{Sachs Electromagnetic Form Factors}
 The  electric $G_E$ and magnetic $G_M$  form factors (known as Sachs form factors) are related to Dirac and Pauli form factors $F_1$ and $F_2$ as follows 
\begin{equation}
\begin{aligned}
G_E (Q^2)\ &= \ F_1 (Q^2)\ -\ \frac{Q^2}{4M^2}\ F_2 (Q^2) \\
G_M (Q^2)\ &= \ F_1(Q^2)\ + \ F_2(Q^2)
\end{aligned}
\end{equation}
Here $M$ is the nucleon mass. Writing $F_1$ \ and $F_2$ \ in terms of $ G_E$\ and $G_M$ we get ($\tau  = \frac{Q^2}{4M^2}$)
\begin{equation}
\begin{aligned}
F_1(Q^2) & = \frac{G_E(Q^2) + \tau G_M(Q^2)}{1 + \tau}, \\
F_2(Q^2) & = \frac{G_M(Q^2) - G_E(Q^2)}{1 + \tau}.
\end{aligned}
\end{equation} 
For the electromagnetic form factors $G_E(Q^2)$ and $G_M(Q^2)$, the data fit from Ref.~\cite{Alberico} were used. 

\subsection{Parametrization of Form Factors Using Generalized Parton Distributions}

\vspace{3mm}

 Generalized parton distribution (GPDs) can be considered as generalization of ordinary parton distributions. The formal definition of GPDs for transversely polarized nucleon but unpolarized quarks is given by
\begin{equation}
\begin{aligned}
\matrixel{p',S'}{\hat{O}_q(x, \textbf{b}_\perp)}{p,S}\ = \ \frac{1}{2 \bar{P}^+} \bar{u}(P',S') \left(\gamma^{+}\ H_q(x,\xi,t) \ + \ i\frac{\sigma^{+ \nu} \Delta_\nu}{2M} E_q(x,\xi,t)  \right)  u(p,S)
\end{aligned}
\end{equation}
Where $\bar{P}^\mu = \frac{1}{2}(P^\mu + P'^\mu)$ represents the average momentum of the target, $\Delta^\mu = P'^\mu - P^\mu$ is the four momentum transfer, $t=\Delta^2$ is the invariant momentum transfer and $\xi = - \frac{\Delta^+}{2 P^+}$ is the change in the longitudinal component of the target momentum and is called the skewness. The nucleons form factors can be decomposed as follows 
\begin{equation}
\begin{aligned}
F^p_i = e_u F^u_i + e_d F^d_i + e_s F^s_i, \ \  \ \
F^n_i = e_u F^d_i + e_d F^u_i + e_s F^s_i
\end{aligned}
\end{equation}
Where $i = 1, 2$ and \ $e_u=\frac{2}{3}, \ e_d = e_s= \frac{-1}{3}$. The Dirac and Pauli flavor form factors at zero skewness are give by the following sum rules
\begin{equation}
\begin{aligned}
F^q_1(t) = \int_{-1}^1 dx \ H^q (x,\xi,t),  \ \  \ \
F^q_2(t) = \int_{-1}^1 dx \ E^q (x,\xi,t)
\end{aligned}
\end{equation}
\noindent The result of integration is independent of $\xi$. Also the integration region can be reduced to $0 < x < 1$ by introducing the non-forward parton densities
\begin{equation}
\begin{aligned}
\mathcal{H}^q (x,t) & = H^q (x,0,t) +  H^q (-x,0,t), \ \ \ \
\mathcal{E}^q (x,t)  = E^q (x,0,t) +  E^q (-x,0,t)
\end{aligned}
\end{equation}
Where $q= u, d$ and $\mathcal{H}^q (x,t)$ reduces to the usual valence quark densities for $t \rightarrow 0$ for the up and down quarks. Now the form factors becomes
\begin{equation}
\begin{aligned}
F^q_1(t)  =  \int_{0}^1 dx \ \mathcal{H}^q (x,t), \ \ \ \ 
F^q_2(t)  =  \int_{0}^1 dx \ \mathcal{E}^q (x,t) 
\end{aligned}
\end{equation}
The magnetic densities satisfies the following normalization conditions
\begin{equation}
\begin{aligned}
\kappa_q  =  \int_{0}^1 dx \ \mathcal{E}^q (x), \ \ \
\kappa_u & = 2 \kappa_p + \kappa_n = + 1.673, \ \ \ 
\kappa_d  = \kappa_p + 2 \kappa_n = - 2.033 \\
F^p_2   (t=0) & = 1.793, \ \ \ \ 
F^n_2  (t=0)  = - 1.913
\end{aligned}
\end{equation}
Following~\cite{GPD2}, the anzats for the $GPDs$ 
\begin{equation}
\begin{aligned}
\mathcal{H}^q (x,t) & = q_v (x) x^{- \alpha' (1-x) t}, \\
\mathcal{E}^q  (x,t) & = \frac{\kappa_q}{N_q} \left (1-x \right)^{\eta_q} q_v x^{- \alpha' (1-x) t},
\end{aligned}
\end{equation}
The normalization constants $N_q$ satisfies
\begin{equation}
N_q = \int _0^1 dx (1-x)^{\eta_q} q_{\nu} (x)
\end{equation}
Where the  unpolarized parton distributions are parametrized as
\begin{equation}
\begin{aligned}
u_\nu (x) & = 0.262 x^{-0.69} (1-x)^{3.50} (1+3.83 x^{0.5} + 37.65 x) \\
d_\nu (x) & = 0.061 x^{-0.65} (1-x)^{4.03} (1+49.05 x^{0.5} + 8.65 x)
\end{aligned}
\end{equation}
The parameters used in this fit are $ \alpha' = 1.105, \eta_u = 1.713, \eta_d = 0.566$.

\end{appendices}

\vspace{5mm}
\bibliographystyle{unsrt}
\bibliography{thesis.bib}

\end{document}